\documentclass[preprint,12pt]{elsarticle}

\usepackage{amssymb}
\usepackage[utf8]{inputenc}
\usepackage{amsmath}
\usepackage{hyperref}
\hypersetup{
     colorlinks = true,
     citecolor  = cyan,  
     linkcolor  = magenta 
}

\newcommand{\revision}[1]{{\color{black}#1}}

\begin{document}

\begin{graphicalabstract}
	\\
	\includegraphics[width=\columnwidth]{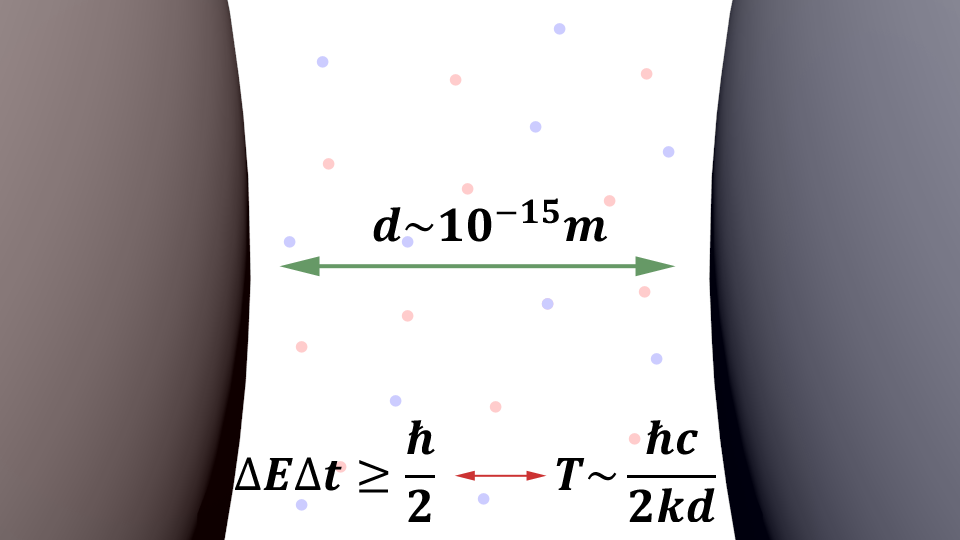}
\end{graphicalabstract}

\begin{highlights}
	\item Links between Uncertainty Relations and Temperature-Distance relations in Casimir physics
	\item Casimir effects  at nuclear length scales
	\item Casimir binding energy, meson mass, and plasmon lifetime
	\item Temperature range relevant for quark-gluon plasma generation
\end{highlights}

\begin{abstract}
	We explore the fundamental idea that there may be a role for the Casimir effect, via an uncertainty relation, in the generation of electron-positron and quark-gluon plasmas. We investigate this concept, reviewing the possible contribution of semi-classical electrodynamics to nuclear interactions, specifically focusing on the Casimir effect at sub-Fermi length scales.  The main result is a temperature distance relation, derived from the time-energy uncertainty relation, which can have observable consequences at these extreme scales.
	{ \revision{From a more general perspective, since the energy-time uncertainty relation appears to be a significant physical quantity, we also provide a brief overview of recent developments in this direction in Sec. 3.2. }}
\end{abstract}


\title{Can An Uncertainty Relation Generate A Plasma?}


\author[label1]{A. Gholamhosseinian}
\author[label2]{R. W. Corkery}
\author[label3]{I. Brevik}
\author[label4,label5]{M. Bostr\"om\corref{cor}}
\ead{mathias.bostrom@ensemble3.eu}

\affiliation[label1]{organization={Freiburg Center for Interactive Materials and Bioinspired Technologies (FIT), University of Freiburg},
	city={Freiburg},
	postcode={79110},
	country={Germany}}


\affiliation[label2]{organization={Department of Materials Physics, Research School of Physics, Australian National University},
	city={Canberra},
	postcode={2601},
	country={Australia}}

\affiliation[label3]{organization={Department of Energy and Process Engineering, Norwegian University of Science and Technology},
	city={Trondheim},
	postcode={NO-7491},
	country={Norway}}

\affiliation[label4]{organization={Centre of Excellence ENSEMBLE3},
	addressline={Wolczynska Str. 133},
	city={Warsaw},
	postcode={01-919},
	country={Poland}}

\affiliation[label5]{organization={Chemical and Biological Systems Simulation Lab, Centre of New Technologies, University of Warsaw},
	addressline={Banacha 2C},
	city={Warsaw},
	postcode={02-097},
	country={Poland}}

\maketitle

\section{Introduction}
Since 1969, Barry Ninham, Adrian Parsegian, and their collaborators have made numerous contributions to the theory and experiments on intermolecular forces\,\cite{ParNin1969,NinhamParsegianWeiss1970,NinPars1970,Pars,Ninhb}. One of the most intriguing ideas is\,\cite{PhysRevA.67.030701,EPJDNinham2014,Ninham_Brevik_Bostrom_2022}) to propose a connection between semi-classical theory (Maxwell's equations, Planck's quantization of light\,$\rightarrow$\, Lifshitz\,\cite{Dzya} and Casimir\,\cite{Casi} interactions) and nuclear interactions.  In this work, we will consider the idea that there is a connection between screened Casimir-Yukawa potentials and quantum mechanical uncertainty. 
Ninham and co-workers\,\cite{Ninham_Brevik_Bostrom_2022} recently revisited the idea that Casimir theory leads to the correct order of magnitude for the meson masses, pion lifetimes, and the Casimir-Yukawa potential contribution to the nuclear binding energy. Here, a temperature-distance uncertainty relation is deduced in a heuristic way, which has implications for the contributing role of Casimir forces in nuclear interactions. This relation is the same as one found earlier\,\cite{Ninham_Brevik_Bostrom_2022} considering a low-temperature expansion of the Casimir interaction between perfect metal surfaces\,\cite{PhysRevA.57.1870}. At an equilibrium point where the first and third terms in the energy expansion cancel out, the dominant chemical potential term corresponds to extremely high temperatures, which may be linked to the generation of an electron-positron plasma\,\cite{landau2013statistical}). We argue that the underlying mechanism for identical results found in two different ways reflects a fundamental relation between the uncertainty principle and Casimir effects.

\section{Materials and Methods}

The interaction between nuclear particles follows a screened Yukawa potential\,\cite{Yukawa1935,Yukawa1955}. This study focuses on a predicted temperature-distance relation, potentially implying quantum electrodynamical contributions to nuclear interactions. From the specific temperature at distance $d$ one can calculate the density of a background electron-positron plasma\,\cite{Ninham_Brevik_Bostrom_2022}. Notably, an electron-positron pair sea can form from available photons in the gap between surfaces through the reaction $e^++e^-\leftrightarrow\gamma$\,\cite{landau2013statistical}. A question too important to dismiss, raised by Ninham, is whether the Casimir effect, under certain conditions, could contribute at nuclear scales. Ninham and Bostr\"om\,\cite{PhysRevA.67.030701} used an approximate approach to compare the magnitude and asymptotic form of this potential with the screened Casimir potential between surfaces in the presence of an electron-positron plasma. Surprisingly, similarities were found that indicated a potential role for the Casimir effect as a non-negligible contribution at nuclear length scales\,\cite{PhysRevA.67.030701,EPJDNinham2014,Ninham_Brevik_Bostrom_2022}.

\section{Results}

\subsection{Caution against particle generation from uncertainty relations}

Many textbooks primarily focused on quantum mechanics discuss how the energy-time uncertainty principle suggests that violation of energy conservation and related particle generation is possible if it obeys $\Delta E \Delta t\leq \hbar/2$\,\cite{Roberts_2020}. 
Recent theoretical work\,\cite{Roberts_2020} has questioned this interpretation of time-energy uncertainty as responsible for the creation of particles. In systems where quantum dynamics are unitary, energy is conserved, and thus uncertainty relations may not be able to explain particle creation processes. The apparent particle creation effects often attributed to $\Delta E \cdot \Delta t \geq \hbar/2$ are arguably better understood as artifacts of approximations in perturbation theory. In quantum systems with creation ($a_i$) and annihilation ($a_i^\dagger$) operators, particle number $N = \sum_i a_i^\dagger a_i$ may not be conserved under the true dynamics if $[N, U_t] \neq 0$ (here $U_t$ is the unitary
propagator). However, arguments have been given\,\cite{Roberts_2020} that this non-conservation does not come directly from uncertainty relations. 
 It was further claimed that "shorter times allow more particle creation." This can be understood through the Mandelstam-Tamm uncertainty principle\,\cite{Roberts_2020},
\begin{equation}
\tau_{\rho(A)} \Delta \rho(H) \geq \frac{\hbar}{2}.
\label{MandelstamTam}
\end{equation}
In this principle, the characteristic time $\tau_{\rho(A)}$ for an observable $A$ to be changed by its standard deviation, is inversely related to the energy spread $\Delta \rho(H)$. When applied to particle number operators, the Mandelstam-Tamm uncertainty principle suggests that rapid changes in the particle number expectation value must be accompanied by large uncertainties in the particle number itself due to the fundamental relationships between observables in quantum mechanics. These limitations of time-energy uncertainty are not related to the formation of electron-positron pairs, that can be created by the quantum field from Casimir effects.
Instead, they indicate that by considering the quantum properties of the electromagnetic field and boundary conditions, particle creation comes from the interaction between the quantum vacuum and the Casimir geometry, not just from uncertainty relations.

\subsection{Generalized thermodynamics derived from the energy-time  uncertainty relation}

As outlined by Uffink\,\cite{Uffink1999} and others\,\cite{Bohr1985,EnergyTemperatureNatureCommun2018}, the uncertainty relations are vital parts of the history of physics.  Some interesting results attributed\,\cite{Uffink1999,EnergyTemperatureNatureCommun2018} to Bohr\,\cite{Bohr1985}, discuss the possibility of a complementary relationship in classical physics, particularly between energy and temperature.  
This energy-temperature uncertainty relation\,\cite{Uffink1999,EnergyTemperatureNatureCommun2018},
\begin{equation}
\Delta \beta \geq \frac{1}{\Delta E}, \label{A}
\end{equation}
 is of considerable interest ($\beta=1/kT$). The relation implies that the better the mean energy of a system is known, the larger the fluctuation in temperature. One might think at first that this is just another form of the Heisenberg uncertainty relation involving positions and momenta, but that cannot be true since the time parameter $t$ is a classical c-number and not an operator. There is definitely a difference, and the connection between these two uncertainty forms may appear delicate. To highlight this point, we will in this subsection give a brief account of how the relation (\ref{A}) can be used to formulate a generalized thermodynamics, of particular interest in micro-systems.  { \revision{As mentioned above, we follow the presentation recently given by  Miller and Anders \cite{EnergyTemperatureNatureCommun2018}, which again is based on a suggestion originally given by Bohr. We now set Boltzmann's constant $k=1$. } }

 The generalization is focused on situations where the interaction is arbitrary, not necessarily weak. Let $S$ refer to the system under study; it is characterized by a temperature $T$ and is in thermal contact with a large reservoir called $R$. For the physically closed system $S\cup R$ we have the Hamiltonian,
 \begin{equation}
 H_{S\cup R}= H_S + H_R + V_{S\cup R},
 \end{equation}
where $H_S$ and $H_{R}$ are the bare Hamiltonians of $S$ and $R$ and $V_{S\cup R}$  the interaction of arbitrary strength. For the total system  $ S\cup R$ we assume the Gibbs form,
\begin{equation}
\pi_{S\cup R}(T)= e^{-\beta H_{S\cup R}}/Z_{S\cup R},
\end{equation}
where  $Z_{S\cup R} = tr_{S\cup R} [\pi_{S\cup R}e^{-\beta H_{S\cup R}}] $ is the partition function.  Corresponding notations are used for the bare systems $S$ and $R$.
{In this paragraph, for clarity, we follow the notation of Miller and Anders, using $U$ to represent internal energy instead of $E$.}
Thus for the subsystem $S$, \revision{$U_S(T) =\partial_\beta \ln Z_S$,} and for the composite system, $U_{S\cup R} = -\partial_\beta \ln Z_{S\cup R}$.

{  \revision{ Because of the coupling, the Hamiltonian $H_S(T)$ will be modified into a new form which Miller and Andrew call an effective Hamiltonian, endowed with an asterisk, thus $H_S^*(T)$. Then the mean energy of the system $S$, called $E_S^*(T)$, can be written as, }}
\begin{equation}
E_S^*(T)= \partial_\beta [\beta H_S^*(T)],
\end{equation}
where,
\begin{equation}
E_S^*(T)= H_S + \partial_\beta[\beta (H_S^*(T)-H_S)].
\end{equation}
The main result of the calculation is that the uncertainty $\Delta \beta_S$ of $S$ can be expressed as,
\begin{equation}
\Delta \beta_S \geq \frac{1}{\sqrt{\Delta U_S^2- Q[\pi_S, E_S^*]}}   \geq \frac{1}{\Delta U_S},
\end{equation}
where $Q$ is a non-negative quantum mechanical entity representing the strong coupling contribution.
The quantity  $Q$ thus leads to a larger lower bound on $\Delta \beta_S$. For the simple case of weak coupling, $Q\rightarrow 0$ and the result of Eq.~(\ref{A}) is recovered.

In conclusion, Bohr's uncertainty relation for energy and temperature plays an independent role, and can even provide a foundation for this generalized form of thermodynamics. Miller and Anders's work on generalized thermodynamics based on Bohr’s original suggestion that there should exist a connection between temperature and energy in thermodynamics ($\Delta \beta\geq 1/\Delta E)$) underlies our main result and will be expanded upon in the next section.

\subsection{Prediction of potential formation of an electron-positron plasma}

A meson lifetime calculation\,\cite{Ninham_Brevik_Bostrom_2022}, using ideas inspired by a Nature paper published by Wick in 1938\,\cite{Wick}, was presented in the unpublished manuscript by Ninham and Pask (see Ref.\,\cite{Ninham_Brevik_Bostrom_2022}). 
It was based on Ninham's published theoretical calculation of the lifetime of an electron plasma that was confirmed by the experiments of Powell and Swanson\,\cite{NinhamPhysRev.145.209}. 
Wick's work was based on the uncertainty principle\,\cite{Heisenberg1927}.  The extremely short duration required for excitations to travel at the light speed ($c$) between two nuclear interfaces (a total distance of \,$d$) was approximated from\,\cite{Wick} $\Delta t\sim d/c$. This timescale is significantly shorter than the meson lifetime.   The meson energy with experimentally known relativistic mass $m_\pi$, $\Delta E\geq m_\pi c^2$, follows from the uncertainty principle\,\cite{Heisenberg1927,LandauLifshitzQM1997}:  $\Delta E\Delta t\geq \hbar/2$. Hence, the expressions for energy and time led Wick to deduce a relationship for nuclear separations with meson mass,
 \begin{equation}
    m_\pi c^2 \times(\frac{d}{c}) \sim \frac{\hbar}{2} \longleftrightarrow m_\pi c^2\sim \frac{\hbar c}{2d}.
     \label{Wickexpanded}
 \end{equation}
As pointed out by Wick, this gives "the distance as the limit up to which virtual transitions can make themselves felt without contradicting the energy principle”\cite{Wick}.
Our main result is based on Bohr's suggestions discussed above that there should exist a connection between temperature and energy in thermodynamics ($\Delta \beta \geq 1/\Delta E$). 
In a similar spirit, notably in the high temperature-short separation limits, we identify the energy with the maximum fluctuations in virtual thermal energy ($kT\sim m_\pi c^2$). These thermal fluctuations {are ultimately} related to the Casimir effect at nuclear distances. Based on these calculations, we propose in the current work a heuristic relation between an effective virtual interaction temperature and nuclear separation,
\begin{equation}
   k T\sim \Delta E\sim \frac{ \hbar c}{2 d}. 
   \label{kTHeisenbergUn1}
\end{equation}
This is a useful relation between uncertainty in energy (i.e., uncertainty in temperature) and uncertainty in time (i.e., particle distance) on the femtometer scale. If this temperature has any physical reality, a natural consequence is that a bath of particles arises from quantum vacuum fluctuations (zero temperature by definition). Here, following Landau and Lifshitz, we analyze it using a known  relationship valid at high temperatures, between temperature and e$^-$-e$^+$ plasma density\,\cite{landau2013statistical}, 
\begin{equation}
\rho=\rho_{-} + \rho_{+}=\frac{3 \zeta(3) k^3 T^3}{ \pi^2\hbar^3 c^3}\sim\frac{3 \zeta(3)}{ 8 \pi^2 d^3}.
\label{eqn:7}
\end{equation}
The last relation in the expression for the induced density of $e^{-}$ and $e^{+}$ arises from Eq.\,(\ref{kTHeisenbergUn1})  (i.e., the relation between temperature and distance between a pair of nuclear particles). The source of the generated energy is fundamentally the uncertainty principle, or equivalently the Casimir force\,\cite{Casi} at nuclear separations. This force is powerful at these extremely short distances.  In an unpublished note by Ninham and Pask from 1969 zero temperature Casimir energy between a pair of perfect metal surfaces was predicted to be large enough to generate a virtual electron-positron plasma.  This in 1969 (when the unpublished note was written) made perfect sense as the photon exchange leading to the Casimir force is virtual.  However, the later work by Mitchell, Ninham, and Richmond indicated that the finite temperature Casimir-Lifshitz force is due to losses of modes in the black body radiation in the gap compared to the fields outside the gap\,\cite{MitchellNinhamRichmond1972_bodyradiationandCasimir}. This leads us to ask: Are the photons and the related electron-positrons in the intervening plasma virtual or real? A related question is the standard interpretation of long-range retardation of dispersion forces\,\cite{PhysRevA.57.1870,PhysRevA.60.2581}. For large enough separations, the zero temperature force depends on the velocity of light. In contrast, for any finite temperature, the long-range force is purely entropic and independent of the velocity of light.

\subsection{Free Energy Expansions if Casimir Interaction Between Perfect Metal Surfaces}

We will next show how the same temperature-distance relation came out from a Casimir free energy expansion\,\cite{PhysRevA.57.1870,Ninham_Brevik_Bostrom_2022}. When considering finite temperature Casimir interaction between perfect metal surfaces in the absence of intervening plasma, Ninham, and Daicic derived the following expansion of the Casimir energy valid at short separations or low temperatures\,\cite{PhysRevA.57.1870},

\begin{equation}
G(d,T)\approx \frac{- \pi^2\ \hbar c}{720d^3}- \frac{\zeta(3) k^3 T^3}{2 \pi\hbar^2 c^2}+\frac{\pi^2 d k^4 T^4}{45\hbar^3 c^3} +..,
\label{BWNapproxFreeEnergy}
\end{equation}
$\zeta(3)$ is a zeta function. 
The first term is the attractive zero temperature Casimir result. The third term is the black body radiation energy between the plates. Ninham and Bostr\"om discussed how it opposes the attractive Casimir term\,\cite{PhysRevA.67.030701}. Ninham assumed\,\cite{PhysRevA.67.030701} that the first and the third terms exactly cancel out at a specific equilibrium. Then the same temperature-distance relation is obtained that we received above from the uncertainty relation. This leaves a dominating chemical potential term.

One real puzzle, in the current context, is that the Casimir-Lifshitz interaction free energy at the femtometer scale is of the same order of magnitude as the nuclear binding energy. This does not imply that semi-classical electrodynamics is an alternative explanation for mesons and nuclear physics. However, it is a strange, intriguing, and suggestive coincidence that deserves further attention. It suggests that the quantum electrodynamics Casimir effect could contribute at very small length scales.
Notably, the temperatures derived from Eq.\,\ref{kTHeisenbergUn1} and Eq.\,\ref{BWNapproxFreeEnergy} (at the Femtometer scale and smaller) are large enough not only to generate an electron-positron plasma but are also in the temperature range relevant for the generation of quark-gluon plasmas.

\revision{
\subsection{Comparison with Experiments}

In order to compare with experiments, we summarize\,\cite{Ninham_Brevik_Bostrom_2022} some numerical results in Table\,\ref{Meson}.  The effective surface area is taken to be $A=\pi r^{2}$ with $r=$ proton radius $\sim 0.8$ fermi. The experimental binding energy per nucleon varies in different atomic nuclei (1.1 MeV in the case of deuterium, while Nickel-62 has 8.8 MeV)\,\cite{Ninham_Brevik_Bostrom_2022}. 
The binding energies presented in the Table\,\ref{Meson} assumes screening as discussed by Ninham, Brevik and Bostr\"om\,\cite{Ninham_Brevik_Bostrom_2022}. However, even the zero temperature Casimir interaction energy provides a value close to the experimental binding energies. Simple estimates for the binding energy for two nuclear particles interacting with a zero temperature unscreened Casimir interaction give a value of the order 10\,MeV. The experimental meson mass is around $264 m_e$\,\cite{Jackson1958,BurchamJobes1995} which is close to our\,\cite{Ninham_Brevik_Bostrom_2022} estimates.
Furthermore, Larin et al.\,\cite{LarinScience_2020} carried out some precise measurements of the lifetime of the $\pi_{0}$ meson. Their weighted average result for the $\pi_{0} \rightarrow \gamma \gamma$ decay width set the lifetime to be of the order $8 \times 10^{-17}$ s. Considering the approximations used, the results are astonishing. All are about the right order of magnitude. However, we do not claim anything more than that the Casimir effects could play a role at nuclear scales. The approximations in our model were discussed in more details by Ninham, Brevik and Bostr\"om\,\cite{Ninham_Brevik_Bostrom_2022}. One important approximation assumes the nucleons to be replaced by conducting plates. We plan to improve this part of the theory and present results for a pair of spherical nuclear particles, both in vacuum and high-density plasmas. The theory for this has been derived by Dr. Larissa Inacio [private communication]. }
\begin{table}
\centering
\begin{tabular}{|c|c|c|c|c|}
 \hline
Separation & Lifetime & Meson Mass & 
Binding Energy & kT \\
\hline
1.0 fermi & $1.61 \times 10^{-17} \mathrm{~s}$ & $491 m_{e}$ & 13.6 MeV & $193 m_{e} c^{2}$ \\ \hline
1.5 fermi & $1.62 \times 10^{-17} \mathrm{~s}$ & $267 m_{e}$ & 4.5 MeV & $128 m_{e} c^{2}$ \\ \hline
2.0 fermi & $1.64 \times 10^{-17} \mathrm{~s}$ & $173 m_{e}$ & 2.0 MeV & $97 m_{e} c^{2}$ \\
\hline
  \hline
\end{tabular}
\caption{\label{Meson} \revision{Some numerical results from the theory for plasmon lifetime, meson mass and binding energy versus separation between a pair of neutrons (or protons).  The data in this table are taken from Ref.\,\cite{Ninham_Brevik_Bostrom_2022}. } }
\end{table}

\section{Discussion}
\label{Discussion}
\par

The assumptions used in the current work are supported by an alternative derivation that explored the same relation between time and temperature $(\Delta t=\hbar/\Delta(kT))$\,\cite{10.1119/1.3194050,Bostrom2025IntJModPhys}. 
One can deduce that, if the ideas explored in the current work hold, distances around one femtometer or smaller correspond to extremely high temperatures ($10^{12}$\,K and higher). These high-temperatures, could potentially generate a quark-gluon plasma at nuclear length scales\,\cite{quarkgluonNature2007}. However, a cautious approach is recommended and thus {encourages} further work to explore this. We leave the potential validation of such ideas to future investigations. In the considered theory, the connection is via particle separation and its link to energy and temperature.  All can be deduced using arguments based on an uncertainty relation between temperature (thermal energy) and distance in Eq\,(\ref{kTHeisenbergUn1}). Future work in the field of Casimir physics will very likely focus on well-established theory and understanding novel experiments at intermolecular distances\,\cite{Lamo1997,Munday2009,RevModPhys.81.1827,SushNP,PhysRevLett.116.240602,RevModPhys.88.045003,SomersGarrettPalmMunday_CasimirTorque,Ser2018,zhao2019stable,esteso2022effect}. It should also be adapted to femtometer length scales and explore the impact of magnetic media in the future\,\cite{Richmond_1971}. 

We urge the scientific community to reflect on the idea that the force in Lifshitz theory at finite temperatures can be due to losses of modes in the black body radiation in the gap compared to those outside the gap\,\cite{MitchellNinhamRichmond1972_bodyradiationandCasimir}. The issue of virtual versus real photons leads to the question of whether virtual versus real electron-positron/quark-gluon particles are generated by the boundaries acting on the electromagnetic field. This is somewhat similar to the different interpretations of long-range effects (retardation) on the dispersion forces\,\cite{PhysRevA.57.1870,PhysRevA.60.2581}. For large particle separations, the Casimir force at zero temperature depends on the velocity of light. However, it is noteworthy that at any finite temperature, the long-range force is dominated by an entropic term that is independent of the speed of light. Notably\,\cite{PhysRevA.60.2581}, the existence of a classical-like potential can be interpreted as a representation of the correspondence principle, where quantum systems show classical behavior at large separations at non-zero temperatures or any separation for very high temperatures. Future insight could be sought from the observation that the same relation can be deduced from an uncertainty relation, and when equating the zero-temperature Casimir term with a repulsive black body radiation term in an expansion of Casimir free energy between two perfect metal surfaces.

\section*{CRediT authorship contribution statement}

{\bf{ Ayda Gholamhosseinian:}} Writing – review -\& editing, Writing – original draft.
{\bf{Robert Corkery:}} Writing – review \& editing, Writing – original draft.
{\bf{Iver Brevik:}} Writing – review \& editing, Writing – original draft.
{\bf{Mathias Bostr\"om:}} Writing – review -\& editing, Writing – original draft, Initiated and supervised the project,  Funding acquisition.

\section*{Declaration of Competing Interest}
The authors declare that they have no known competing financial interests or personal relationships that could have appeared to influence the work reported in this paper.

\section*{Acknowledgement}
This research is part of the project No. 2022/47/P/ST3/01236 co-funded by the National Science Centre and the European Union’s Horizon 2020 research and innovation programme under the Marie Skłodowska-Curie grant agreement No. 945339. Institutional and infrastructural support for the ENSEMBLE3 Centre of Excellence was provided through the ENSEMBLE3 project (MAB/2020/14) delivered within the Foundation for Polish Science International Research Agenda Programme and co-financed by the European Regional Development Fund and the Horizon 2020 Teaming for Excellence initiative (Grant Agreement No. 857543), as well as the Ministry of Education and Science initiative “Support for Centres of Excellence in Poland under Horizon 2020”.






\bibliographystyle{elsarticle-num} 
\bibliography{Heisenberg}






\end{document}